\documentclass[prb,twocolumn,superscriptaddress,showpacs,unsortedaddress]{revtex4}
\usepackage{graphicx}
\usepackage{dcolumn}
\usepackage{amsmath}
\usepackage{bm}
\usepackage{color}
\usepackage{SIunits}

\begin{document}

\title{A critical re-examination of resonant soft x-ray Bragg forbidden reflections in magnetite.}

\author{S.B. Wilkins} 

\affiliation{Brookhaven National Laboratory, Condensed Matter Physics
\& Materials Science Department, Upton, New York, 11973-5000, USA}

\author{S. Di Matteo} 

\affiliation{\'Equipe de Physique des Surfaces et Interfaces, Institut de Physique de Rennes UMR UR1-CNRS 6251, Universit\'e de Rennes 1, F-35042 Rennes Cedex, France}

\author{T.A.W. Beale} 

\affiliation{Department of Physics, University of Durham, South Rd., Durham, DH1 3LE, UK}

\author{Y. Joly} 

\affiliation{Institut N\'eel, CNRS and Universit\'e Joseph Fourier, B.P. 166, F-38042 Grenoble Cedex 09, France}

\author{C. Mazzoli}

\affiliation{European Synchrotron Radiation Facility, BP 220, F-38043
Grenoble Cedex 9, France}

\author{P.D. Hatton} 

\affiliation{Department of Physics, University of Durham, South Rd., Durham, DH1 3LE, UK}

\author{P. Bencok}
\author{F. Yakhou}

\affiliation{European Synchrotron Radiation Facility, BP 220, F-38043
Grenoble Cedex 9, France}

\author{V.A.M. Brabers}
\affiliation{Department of Physics, Eindhoven University of Technology, NL-5600, MB Eindhoven, The Netherlands}

\begin{abstract}
Magnetite, Fe$_3$O$_4$, displays a highly complex low temperature crystal structure that may be charge and orbitally ordered. Many of the recent experimental claims of such ordering rely on resonant soft x-ray diffraction at the oxygen K and iron L edges. We have re-examined this system and undertaken soft x-ray diffraction experiments on a high-quality single crystal. Contrary to previous claims in the literature, we show that the intensity observed at the Bragg forbidden (00$\frac{1}{2}$)$_c$ reflection can be explained purely in terms of the low-temperature structural displacements around the resonant atoms. This does not necessarily mean that magnetite is not charge or orbitally ordered, but rather that the present sensitivity of resonant soft x-ray experiments does not allow conclusive demonstration of such ordering.  
\end{abstract}

\date{\today}
\maketitle


In many transition-metal oxides, the spatial localization of electrons on certain sites, so-called charge ordering, has been used to explain some of their more intriguing ground-state properties. For example, charge ordering has been invoked to describe phase transitions in some magnetoresistive manganites,\cite{Mori:1998p473} and the dynamic fluctuations of charge-ordered stripes\cite{Tranquada:1995p479} have been proposed as a mechanism of high temperature superconductivity\cite{Salkola:1996p480}.  Magnetite, Fe$_3$O$_4$, was the first material in which such a charge ordering transition was proposed, in connection with the metal-insulator transition discovered by Verwey\cite{Verwey:1947p483}, and it has long been interpreted as the classic example of mixed-valence compound with formula unit Fe$^{3+}$[Fe$^{2+}$Fe$^{3+}$]O$_4$ [Refs.~\onlinecite{Verwey:1947p483,Coey:2004p482}]. In this interpretation, magnetite, which at room temperature crystallises into the cubic inverted spinel structure AB$_2$O$_4$, with space group {\it Fd$\overline3$m}, has formally Fe$^{3+}$ ions at the tetrahedral A sites and formally Fe$^{2+}$ and Fe$^{3+}$ ions at the octahedral B sites. Unfortunately, this simple picture is deceptive as the crystal structure below T$_V$ is extraordinarily complicated: the most recently reported structure\cite{Wright:2001p86} consists of a complex {\it P2/c} monoclinic cell containing 56 atoms in which the A and B Fe ions are split between two and six inequivalent sites respectively.  Based on the {\it P2/c} structure, LSDA+U band structure calculations have reported both charge ordering (CO) (0.16 electrons) and an associated t$_{2g}$ orbital ordering (OO) on the octahedral sublattice \cite{Leonov:2004p678,Jeng:2004p489}. 
Seemingly arguing against this, however, are the results of resonant x-ray scattering (RXS) experiments at the iron K-edge, have been interpreted as providing evidence either against any charge ordering \cite{Garcia:2000p487,Subias:2004p488,Garcia:2004p485}, or in favor of a  0.12 electrons charge disproportionation \cite{Nazarenko:2006p427} between the formally Fe$^{2+}$ and Fe$^{3+}$ sites than predicted in Refs. \cite{Leonov:2004p678,Jeng:2004p489}. Very recently two further papers have appeared in which soft-x-ray diffraction measurements at the O K-edge\cite{Huang:2006p80} and the Fe L-edges\cite{Schlappa:2007p556} were interpreted as providing evidence for both charge and orbital ordering.
We not that in the monoclinic cell the iron B sites are no longer equivalent by symmetry and there is therefore no requirement that they have the same charge density surrounding the atomic site. It is thus most likely that they will not have the same charge density. The question is therefore: what is the smallest charge difference, about which one would reasonably claimed that the material is charge ordered?

Resonant x-ray scattering occurs when a photon excites a core electron into an excited state and is subsequently re-emitted when the electron and core hole combine\cite{HANNON:1988p46}. On resonance the x-ray scattering amplitude is anisotropic and is sensitive to the anisotropic charge distribution of the resonating ion. The anisotropic charge distribution can be intrinsic to the scattering ion due to orbital occupation or can be intrinsic to the lattice as in the case of Templeton-Templeton scattering\cite{Templeton:1982p676,Templeton:1980p675,Dmitrienko:1983p677}. The characteristic of Templeton-Templeton scattering is that a reflection which is Bragg-forbidden because of a compound symmetry operation, such as a glide plane or screw axis, becomes allowed when the incident photon energy is tuned to a resonance. On resonance the x-rays are sensitive to the quadrupolar term in the charge distribution, of the resonating atom, $Q$, and the difference between the two electric quadrupole moments, related by the symmetry operation, sum to zero and a resonant peak is observed arising from the crystal structure.

We have chosen to revisit magnetite and report here on resonant soft x-ray experiments that confirm the resonant enhancement of the (00$\frac{1}{2}$)$_c$ reflection at both the oxygen K- and iron L$_3$ edges. However, we have carried out a careful analysis of this superlattice reflection focussing on a detailed investigation of the effects of the distorted crystal structure below the Verwey transition, \emph{without} invoking any charge or orbital ordering. We find that we can model our data well by considering only Templeton-Templeton scattering arising due to the structural distortions below the Verwey transition, without the need to resort to charge order or orbital order. This is contrary to the claims of Refs.~\onlinecite{Huang:2006p80,Schlappa:2007p556}. 

The experiments were conducted on high-quality synthetic magnetite crystals prepared in an arc-image furnace using the floating-zone technique. The purity of the sample was verified by heat capacity measurements, which gave a maximum heat capacity value of 120.6~K and an entropy change of 5.77~$\mathrm{J}\mathrm{K}^{-1}$ at the Verwey transition. These results give for Fe$_{3-\delta}$O$_4$ a value $\delta= 0.0002$ showing that the stochiometry of our sample is very close to the ideal case.
Soft x-ray diffraction experiments were conducted on the ID08 beamline at the ESRF, Grenoble, France. In what follows we index the sample in the approximate low temperature {\it Pmca} orthorhombic structure (No. 57) with lattice parameters $a = 5.944$~\AA, $b = 5.925$~\AA\ and $c = 16.775$~\AA \cite{Wright:2001p86}. This structure is related to the {\it P2/c} (No. 13) structure by only a slight monoclinic distortion ($\beta = 90.2363^\circ$). In this orthorhombic setting, the cubic (00$\frac{1}{2}$)$_c$ reflection becomes the orthorhombic (001)$_o$. The sample of Fe$_3$O$_4$ was cut with a $[001]_o$ surface normal and polished with 0.1~\micro\meter\ diamond paste. It was then mounted on a SmCo magnet, providing a field at the sample surface of $\approx0.3$~T, parallel to the surface normal. This field defines a unique $c$-axis so that on cooling through the transition the number of crystallographic domains are minimized.  The sample was then to a base temperature of 30~K and a resonant signal was observed at the (001)$_o$ position in reciprocal space in the vicinity of the iron $L_{2,3}$ and oxygen K-edges. 

\begin{figure}
\includegraphics[width=0.8\columnwidth]{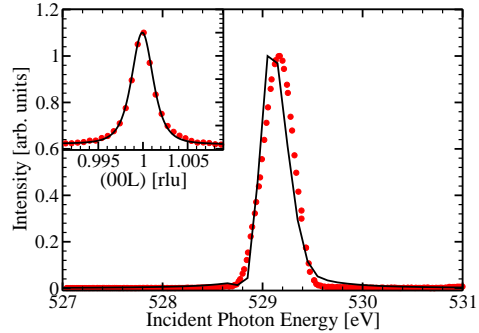}
\caption{(Color online) The incident photon-energy-dependence of the (001)$_o$ reflection in Fe$_3$O$_4$ close to the oxygen K-edge (red circles). The solid black line represents {\it ab-initio} calculations of the scattered intensity assuming structural distortions and no charge or orbital order. In the insert, a scan along the [001]$_o$ direction through the (001)$_o$ reflection is shown (red circles); the solid line is a fit to a Lorentzian squared lineshape.} \label{fig:okedge}
\end{figure}

\begin{figure}
\includegraphics[width=0.8\columnwidth]{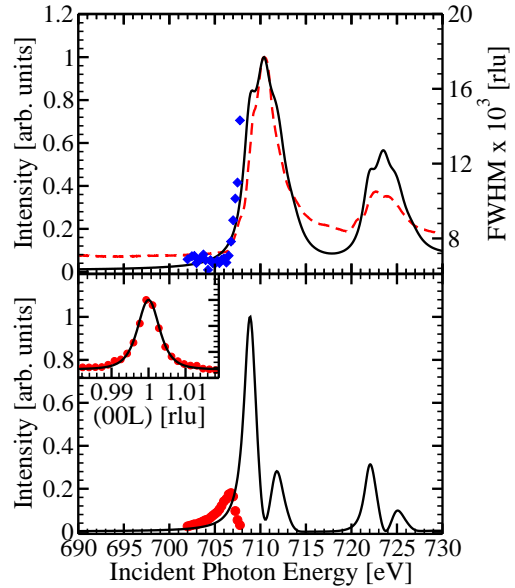}
\caption{(Color online) Top panel: Measured total electron yield (x-ray absorption spectrum) at 27~K  (red dashed line) and calculated absorption spectrum (solid black line) through the Fe~L$_{2,3}$ edges. Bottom panel: incident photon energy dependence of the integrated intensity of the (001)$_o$ reflection close to the Fe~L-edges (red circles). The solid black line shows {\it ab-initio} calculations of the resonant scattering (see text). The inset represents a scan along the [001]$_o$ direction at an energy corresponding to  the maximum in the signal. The variation of the longitudinal width of the (001)$_o$ reflection with incident energy is included in the top panel for comparison with the experimental and calculated absorption (blue diamonds).} \label{figtheo2}
\end{figure}

\begin{figure}
\includegraphics[width=0.8\columnwidth]{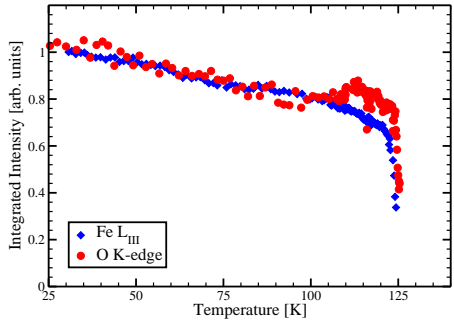}
\caption{(Color online) Temperature dependence of the (001) reflection of Fe$_3$O$_4$ at the O K-edge (red circles) and the iron L$_{3}$ edge (blue diamonds) .} \label{fig:tdep}
\end{figure}

Figure~\ref{fig:okedge} shows an incident photon energy scan at constant wavevector of (001)$_o$ as the  is tuned in the vicinity of the oxygen K-edge. The scattering is observed to peak at  529.1~eV, about 10~eV below the main oxygen K-edge. The insert shows a scan along the [001]$_o$ direction through the (001)$_o$ reflection at $E_i = 529.1$~eV, with a fit to a Lorentzian squared lineshape. The correlation length obtained from this fit is $> 3000$~\AA. This represents a lower bound on the penetration depth of the x-rays and thus this value indicates that the resonant signal is not surface sensitive even at the maximum of the resonance. The (001)$_o$ reflection was also visible in the vicinity of the iron L$_3$ edge (Fig.~\ref{figtheo2}). The bottom panel of Fig.~\ref{figtheo2} shows the incident photon energy dependence of the integrated intensity of the (001)$_o$ reflection. The experimental signal is only visible at the L$_{3}$ threshold, with a maximum at an energy of 706.5~eV and is found to be suppressed above 708~eV. Such behavior arises from the very large self-absorption caused by the strong Fe L$_3$ resonance, leading to a total loss in the observed signal. The width of the diffraction peak as a function of energy as shown in the top panel of Fig.~\ref{figtheo2} confirms this. The peak width is broader than that found at the oxygen K-edge and tracks the calculated absorption (dashed line) indicating that the change in width arrises from the increased absorption and consequently reduced penetration depth. Finally, Fig.~\ref{fig:tdep} shows the temperature dependence of the integrated intensity of the (001)$_o$ reflection measured at both the iron L$_3$ and oxygen K edges. The data were collected by performing rocking scans of the sample angle, $\theta$, at each temperature. The signal at both edges was found to be virtually constant up until a temperature of $\approx 125$~K above which no intensity is observed. 

We now turn to our resonant scattering simulation. We have used the FDMNES program\cite{Joly:2001p293} in the multiple scattering mode. *** The results of these sumulations are shown in Fig.~\ref{fig:okedge} and Fig.~\ref{figtheo2} for the oxygen K and iron L edges respectively. In order to calibrate to the experimentally obtained data with the FDMNES simulations, the calculated absorption was compared with the sample absorption measured by total electron yield at the oxygen K-edge. Our simulation reproduces well the main experimental features, including the energy gap of about 10 eV between the RXS signal and the main oxygen absorption edge, as well as the energy width of the peak. 

In this specific case, there are eight in-equivalent oxygen sites of $4d$ Wyckoff symmetry in the {\it Pmca} space group. Considering only the oxygen atoms which dominate at this energy, the structure factor of the (001)$_o$ reflection is: 
\begin{equation}
S_{(001)_o}=\sum_{j=1\ldots8} 2f_j(1 - {\hat {m}}_y)\cos(2\pi w_j) ,
\end{equation}
where $f_j$ is the atomic scattering amplitude, ($j=1\ldots8$ labels the inequivalent sites), $w_j$ is the fractional coordinate of the $j$th oxygen atom in the $c$ direction and ${\hat {m}}_y$ is the mirror plane in the $b$ direction of the {\it Pmca} setting \cite{Wright:2001p86}. Using the local mirror symmetry ${\hat {m}}_x$ of the $4d$ sites, we find that $f_j \propto {{Q}}_{yz}^{j}$, the electric quadrupole matrix element. In the monoclinic {\it P2/c} setting, the ${\hat {m}}_x$ symmetry is lost and a further contribution $f_j \propto {{Q}}_{xy}^{j}$ appears. This can be shown to be negligible, since it is proportional to the small angular distortion, $\beta\neq 90^\circ$, from the orthorhombic {\it Pmca} structure. Upon evaluating the structure factor $S$ we can conclude that almost all the scattered intensity comes from the sum of the quadrupoles $Q_{yz}$ at oxygen sites O$_1$ and O$_2$ only. That is $S_{(001)_o} \approx {{Q}}_{yz}^{O_1} + {{Q}}_{yz}^{O_2}$
In the high-temperature phase $Q_{yz}^{O_1}=-Q_{yz}^{O_2}$, due to the ${\hat{C}}_{2z}$ screw axis of the high temperature {\it Fd${\overline{3}}$m }space group, and therefore their sum is zero. Below the Verwey transition, the unequal atomic displacements of the O$_1$ and O$_2$ oxygen sites from their high temperature positions, gives a finite signal even if $Q_{yz}^{O_1} = -Q_{yz}^{O_2}$. However, this signal is tiny because of the very small displacement, and has an expected amplitude of $\approx 10^{-4}\times|Q_{yz}^{O_1}|$. In contrast, a much bigger amplitude might be expected if the surrounding iron tetrahedra are distorted making $Q_{yz}^{O_1}\neq -Q_{yz}^{O_2}$.

To investigate which of these contributes to be the most significant we have performed several numerical calculations in which the tetrahedral or octahedrally coordinated iron atoms, and/or the oxygens were in turn placed in their high-temperature positions, with the rest of the cluster held in their low-temperature positions. By this method, we found that the main contribution to the signal comes from the O$_2$ position ($\sim 70 \%$ of the total): in fact $Q_{yz}^{O_1}$ does not change much when the iron sites move from the high-temperature to the low-temperature positions, while $Q_{yz}^{O_2}$ varies by about 100 $\%$. This dominant change is due in particular to the displacement of the octahedral iron atoms surrounding the O$_2$ cite, The iron sites belong to the Fe $B3$ sites, in the notation of Ref.~\onlinecite{Wright:2002p290} and undergo the strongest distortion when passing from the high-temperature to the low-temperature phase: the Fe$_{B3}$-O$_2$ distance changes from $2.06$ \AA\ to $1.96$ \AA, about a $5\%$ contraction. Therefore the O K-edge signal is mainly determined by the hybridization of $2p$ oxygen orbitals at O$_2$ sites with $3d$ iron orbitals belonging to octahedral Fe $B3$ sites.

The fact that we can explain the signal through these atomic displacements contradicts the interpretation of the signal at (001)$_o$ in Ref.~\onlinecite{Huang:2006p80}, that explicitly excluded a structural origin to the (001)$_o$ reflection at the oxygen K-edge. In particular, the arguments made by the authors of Ref.~\onlinecite{Huang:2006p80}, for the assignment of their signal to charge and orbital order due to its polarization dependence, are reproduced in our calculations based solely on structural distortions. This makes clear that invoking charge or orbital ordering to explain the detection of this superlattice reflection is unnecessary, and potentially misleading.

This same procedure was then used to evaluate the resonant signal of the (001)$_o$ reflection at Fe L$_{2,3d}$-edges, as shown in Fig.~\ref{figtheo2}. The calibration between our experimental data and the FDMNES simulation  was set by comparison of the calculated absorption with the absorption as measured by total electron yield. 

We can repeat the same analysis as performed at the oxygen K-edge for the data at the iron L$_{2,3}$ edges. The results of the FDMNES simulation are shown in Figure~\ref{figtheo2}. In the top panel a comparison between the measured and calculated absorption are shown, while the bottom panel shows the comparison between the integrated intensity of the (001)$_o$ reflection as a function of incident photon energy and our simulation. 

In the {\it Pmca} setting there are 6 inequivalent groups of iron atoms, with two groups of tetrahedral iron-sites ($A1$ and $A2$, following the notation of Ref. [\onlinecite{Wright:2002p290}]), and 4 groups of octahedral iron-sites ($B1$, $B2$, $B3$, and $B4$), each group containing 4 iron atoms.
$A1$, $A2$, $B3$, and $B4$ sites have a local ${\hat{m}}_x$-symmetry ($4d$ Wyckoff site), so that the same considerations discussed above for the structure factor at the oxygen K edge are still valid. The only signal that can be measured for these ions is due to the ${\hat {Q}}_{yz}^{j}$, this time projected on the corresponding iron-sites $j$. The $B2$ site has a ${\hat{C}}_{2y}$ local symmetry ($4c$ Wyckoff site) and the two groups of two ions that contribute in antiphase at the (001)$_o$ are related by inversion symmetry, so their total contribution equal to zero. Finally the $B1$ sites, with local inversion symmetry ($4b$ Wyckoff site), also contribute with the quadrupole component ${\hat {Q}}_{yz}^{j}$. 
When we numerically compare the separate contributions of the $A1$, $A2$, $B1$, $B3$ and $B4$ sites, we find that the intensity from the $B1$ site is smaller by a factor of 500 relative to the contribution from the $B4$ site. We can conclude therefore, that the dominant contribution of the intense low-energy peak seen in the calculations at 708.5~eV is from the iron atoms sitting at the $B4$ sites. The tetrahedral $A$ sites and the octahedral $B3$ site contribute mainly to the smaller shoulder at higher energy, and the fact that they have opposite amplitudes gives rise to the local minimum at 710~eV. Therefore all the experimentally detected signal comes from the $B4$ sites (nominally Fe$^{2+}$). This is in keeping with some previous results obtained with rather different approaches which concluded that the low-energy part of L$_3$ spectrum (around 707~eV) is mainly determined by the $t_{2g}$ states of the nominally divalent iron ions \cite{Leonov:2004p678,Kuiper:1997p298}. 
However there is an important difference between the work presented here and atomic-multiplet-based calculations such as those reported in Refs. \onlinecite{Schlappa:2007p556,Kuiper:1997p298}. By focussing on the orbital occupancy in a ionic model, these latter authors have neglected the important structural differences that exist between sites like $B1$ and $B4$ (that are otherwise equivalent where formal charge is concerned, both being formally Fe$^{2+}$). The fact that we calculated their relative contribution to the total signal in the ratio 1:500 proves such a difference.

The similar temperature dependence shown in Fig.~\ref{fig:tdep} can now be explained as a natural consequence of the signals at the oxygen K-edge and iron L-edges both measuring the same order parameter. As discussed above we argue that this order parameter is structural distortions associated with the structural phase transition from the {\it Fd$\overline3$m} high temperature structure to the low temperature {\it P2/c} structure and \emph{not} that of any charge or orbital order.

In conclusion, we have shown that the (001)$_o$ reflection of Fe$_3$O$_4$ is sensitive to the local displacements around the resonant ion at both the oxygen K and iron L edges. The electronic anisotropy arising from the crystal distortions are sufficient to explain the origin of the scattered signals, and any invocation of charge ordering and/or orbital ordering is not necessary to reproduce the data. At the oxygen K-edge, the signal is determined by the hybridization of O $2p$ orbitals of the four O$_2$ atoms with the $B3$ Fe $3d$ orbitals (nominally Fe$^{3+}$). At the iron L$_3$ edge, in contrast, the resonant x-ray scattering is mainly sensitive to the contribution of Fe $3d$ orbitals from $B4$ sites (nominally Fe$^{2+}$). 

Work at Brookhaven was supported by the U.S. Department of Energy under contract DE-AC02-98CH1-886. SBW would like to thank J.P. Hill for critical reading of the manuscript and S.R. Bland for helpful discussions. 

\bibliography{fe3o4_soft_sbw}

\end{document}